\documentclass[useAMS,usenatbib,fleqn,usedcolumn]{mnras}

\usepackage[british]{babel}             % British English hyphenation
\usepackage{graphicx}                   % Including figures

\usepackage{bm}
\usepackage{amsmath}
\usepackage{amssymb}
\newcommand{\unit}[1]{\ensuremath{\,\mathrm{#1}}}
\newcommand{\diff}{\ensuremath{\mathrm{d}}}

\newcommand{\NV}{N~\textsc{v}}
\newcommand{\CIV}{C~\textsc{iv}}
\newcommand{\SiIV}{Si~\textsc{iv}}
\newcommand{\angstrom}{\textup{\AA}}

\title[Accretion and chemical evolution]{Chemical enrichment of the planet forming region as probed by accretion}
\author[Booth \& Clarke]{Richard A. Booth\thanks{E-mail: rab200@ast.cam.ac.uk}$^1$ and Cathie J. Clarke$^1$ \\
$^{1}$Institute of Astronomy, University of Cambridge, Madingley Road, Cambridge, CB3 0HA, United Kingdom \\
}

\begin{document}
\date{Accepted 2017 September 7. Received 2017 September 7; in original form 2017 June 17}
\pagerange{\pageref{firstpage}--\pageref{lastpage}} \pubyear{2017}
\maketitle

\label{firstpage}

\begin{abstract}
The chemical conditions in the planet forming regions of protoplanetary discs remain difficult to observe directly. Gas accreting from the disc on to the star provides a way to measure the elemental abundances because even refractory species are in an atomic gaseous form. Here we compare the abundance ratios derived from UV lines probing T Tauri accretion streams to simple models of disc evolution. Although the interpretation of line ratios in terms of abundances is highly uncertain, discs with large cavities in mm images tend to have lower Si emission. Since this can naturally be explained by the suppressed accretion of dust, this suggests that abundance variations are at least partially responsible for the variations seen in the line ratios. Our models of disc evolution due to grain growth, radial drift and the flux of volatile species carried as ices on grain surfaces, give rise to a partial sorting of the atomic species based on the volatility of their dominant molecular carriers. This arises because volatiles are left behind at their snow lines while the grains continue to drift. We show that this reproduces the main features seen in the accretion line ratio data, such as C/N ratios that are a few times solar and the correlation between the Si to volatile ratio with mm-flux. We highlight the fact that developing a more robust linkage between line ratios and abundance ratios and acquiring data for larger samples has the potential to cast considerable light on the chemical history of protoplanetary discs.  
\end{abstract}

\begin{keywords}
protoplanetary discs ---  accretion, accretion discs --- stars: abundances
\end{keywords}

\section{Introduction}

 Recent years have seen considerable progress in characterising the chemical composition of protoplanetary discs via analyses of molecular tracers
 \citep[e.g.][]{Bergin2013,Favre2013,McClure2016}.
  These studies probe the mass reservoir at relatively large radii (many tens of A.U.) which is where 
the bulk of the disc mass resides; additionally, in recent years
interferometers -- in particular ALMA --  have provided the opportunity to also provide some spatial information on
chemical tracers on these scales 
\citep{Guilloteau2016,Miotello2016,Schwarz2016}.
There are however some drawbacks to such
studies. Firstly, many chemical species are in the form of molecular ices at
such radii or else subject to photodissociation in irradiated surface layers.
Thus atomic/molecular line data needs to be interpreted via complex thermochemical
models \citep[e.g.][]{Miotello2016}. Secondly, the lack of observed planets at such radii \citep{Biller2013,Vigan2017}
suggests that such outer regions may not be representative of chemical
conditions in the planet forming regions of discs.

 A complementary window into disc chemical composition is provided by studies
conducted at the opposite extreme, i.e. those that probe material that
accretes from the disc inner edge on to the star. At such radii even refractory
species are in gaseous, atomic form which makes it easier to infer elemental abundances
from spectroscopic data. This offers an opportunity to study how
material in the inner (planet forming) region of the disc has 
been processed or filtered by its inward transport through the disc. 

The radiative nature of Herbig Ae/Be stars means that accreted material leaves an imprint
on photospheric abundances. \citet{Kama2015} used abundances obtained from high-resolution optical spectra by \citet{Acke2004}  and \citet{Folsom2012} to demonstrate that the abundance of
refractory elements (Mg, Fe, Si) is depleted in stars showing evidence for inner cavities in
their dust discs. While this is perhaps unsurprising, given that these refractories
are the bulk constituent of the dust that controls the disc continuum emission, it is
of particular interest that there is no comparable depletion of so-called volatiles (C, N, O etc.).
This implies that these  species are not predominantly found in a solid form 
(refractory grains/ices) that share in the processes that control the retention of
dust in the disc.

 T Tauri stars are however fully convective and the short convective turnover time
{\citep {Drake2005A} means that accreted material is rapidly mixed below
the stellar photosphere. Photospheric abundances in T Tauri stars thus represent
the bulk composition of the star, accrued during the entire formation/disc accretion
process. The current accretion onto the star is however observationally accessible via ultraviolet/X-ray spectroscopy.
For example, \citet{Herczeg2002} found an empirical connection between FUV emission and the volatile abundance of the accretion flow.  \citet{Kastner2002} and \citet{Drake2005A} then  showed an enhanced level of Ne/O in  X-ray spectroscopy
of TW Hydra when compared either with the Classical T Tauri star BP Tau or
with emission from  stellar coronae in non-accreting post Tauri stars \citep{Drake2005N}.
This is notable because TW Hydra 
 is well known to possess an inner cavity in  its dust emission \citep{Calvet2002,Hughes2007,Andrews2016} in contrast to BP Tau which is an accreting star that is
apparently devoid of structure in its circumstellar dust (although it is possible that if imaged at the same high resolution to TW Hydra then some structure would be found). 
\citeauthor{Drake2005A} tentatively
suggested that this implies the preferential retention of oxygen (in the form of
ices) within the dusty structures of TW Hydra. Similarly, \citet{Kastner2002} and \citet{Stelzer2004} found depletion of Fe and O in TW Hydra, while \citet{Gunther2006} found a high Ne/O ratio for the  binary V4046 Sgr and \citet{France2010} found evidence for the depletion of refractories in the brown dwarf 2M1207.

 The only comparative dataset on chemical compositions at the star-disc interface in
T Tauri stars is contained in  the ultraviolet spectroscopic study of \citet{Ardila2013}. 
These authors studied the resonant doublets of \NV{}, \SiIV{} and \CIV{} in a large sample of T Tauri stars including both accreting 
and non-accreting systems. They concluded that while \NV{} and \CIV{} emission is
rather closely correlated, there is considerable scatter in the ratio of
\SiIV{} to \CIV{} emission. They however steered away from the conclusion that
this scatter represented an evolutionary effect since they found no
evidence that the Si to C ratio correlated with accretion rate or the designation
of the  system as a transition disc.

 In this paper we re-examine this dataset from the perspective of a modern
understanding of chemical processing in discs. In \autoref{Sec:Obs} we review the data
and explore possible connections with disc evolutionary status using a classification
scheme based on more recent imaging data, in particular examining
the connection between abundance ratios and dust continuum emission at
mm wavelengths. In \autoref{Sec:Model} we briefly describe our model for the chemical evolution driven by the growth and transfer of grains
in protoplanetary discs. We apply this to the issue of the signature
of such processes in terms of abundance ratios in accreting T Tauri stars.

\section{Observations}
\label{Sec:Obs}

\begin{table*}
\caption{Observational data used in this work.}
\label{Tab:Data}
\begin{tabular}{l c rrr cc l}
\hline
Name & $F_{\rm 1.3\,mm}$$^1$ & Si/N$^2$ & Si/C$^2$ & C/N$^2$ & Spec. Type & Outflow$^{5}$ & Ref. for spec. type and $F_{\rm 1.3\,mm}$\\
\hline
\multicolumn{8}{c}{Classical T Tauri Stars -- Full Discs}\\
\hline
AA Tau & 88 & $<1.8$ & $<0.07$ & 25.0 & K7 & No & \citet{Donati2010,Andrews2005}\\
AK Sco & 36 & 5.6 & 0.56 & 10.0 & F5 & Yes & \citet{Andersen1989,Jensen1996}\\
BP Tau & 47 & 2.5 & 0.15 & 16.7 & K7 & No & \citet{Johns-Krull1999,Andrews2005}\\
CV Cha & 18$^3$ & 10.3 & 0.62 & 16.7 & G8 & No & \citet{Gunther2007,Belloche2011}\\
CY Tau & 133 & $<5.9$ & $<0.88$ & 6.7 & M1 & No & \citet{Hartmann1998,Guilloteau2011}\\
DE Tau & 36 & $<2.0$ & $<0.12$ & 16.7 & M0 & No & \citet{Furlan2009,Andrews2005}\\
DK Tau & 35 & 7.4 & 0.37 & 20.0 & K7 & Yes & \citet{Furlan2009,Harris2012}\\
DN Tau & 84 & 1.0 & 0.10 & 10.0 & M0 & No & \citet{Furlan2009,Andrews2005}\\
DR Tau & 159 & 10.0 & 0.31 & 33.3 & K7 & No & \citet{Petrov2011,Andrews2005}\\
DS Tau & 25 & $<1.4$ & $<0.07$ & 20.0 & K5 & No & \citet{Muzerolle1998,Andrews2005}\\
EP Cha & 83$^4$ & 1.3 & 0.10 & 12.5 & K4 & No & \citet{Mamajek1999,Riviere-Marichalar2015}\\
ET Cha & 7$^3$ & $<1.6$ & $<0.28$ & 5.6 & M2 & No & \citet{Lawson2002,Woitke2011}\\
%HN Tau A & 13 & 3.8 & 0.88 & 4.3 & K5 & Yes & \citet{Furlan2009,Akeson2014}\\
IP Tau & 16 & 3.6 & 0.18 & 20.0 & M0 & No & \citet{Furlan2009,Andrews2005}\\
MP Mus & 84 & 1.6 & 0.14 & 11.1 & K1 & No & \citet{Torres2006,Grafe2013}\\
RU Lup & 190$^3$ & 15.5 & 1.00 & 16.7 & K7 & No & \citet{Herczeg2005,Ansdell2016}\\
%RW Aur & 42 & 11.0 & 1.56 & 7.1 & K4 & Yes & \citet{Johns-Krull1999,Andrews2005}\\
SU Aur & 23 & 4.3 & 0.30 & 14.3 & G1 & No & \citet{Furlan2009,Pietu2014}\\
T Tau N & 280 & 3.6 & 0.36 & 10.0 & K1 & Yes & \citet{Walter2003,Andrews2005}\\
V1190 Sco & 5$^4$ & $<0.7$ & $<0.08$ & 9.1 & K0 & No & \citet{Hughes1994,Bustamante2015}\\

\hline
\multicolumn{8}{c}{Classical T Tauri Stars -- Cavity Sources}\\
\hline
CS Cha & 167 & $<0.5$ & $<0.15$ & 2.9 & K6 & No & \citet{Luhman2004,Ubach2012}\\
DM Tau & 109 & $<0.7$ & $<0.09$ & 7.7 & M1 & No & \citet{Espaillat2010,Andrews2005}\\
GM Aur & 253 & $<0.6$ & $<0.07$ & 9.1 & K5 & No & \citet{Espaillat2010,Andrews2005}\\
LkCa 15 & 167 & 3.9 & 0.25 & 16.7 & K3 & No & \citet{Espaillat2010,Andrews2005}\\
TW Hya & 140 & 0.3 & 0.04 & 8.3 & K6 & No & \citet{Torres2006,Weintraub1989}\\
UX Tau A & 63 & 1.4 & 0.27 & 5.3 & K2 & No & \citet{Herbig1977,Andrews2005}\\
V4046 Sgr & 77 & $<0.1$ & $<0.04$ & 1.4 & K5 & No & \citet{Quast2000,Rosenfeld2013}\\

\hline
\multicolumn{8}{c}{Weak T Tauri Stars}\\
\hline
EG Cha    &  & 2.0 & 0.69 & 2.9  & K4 & & \citet{Mamajek1999} \\
TWA 7     &  & 1.5 & 0.43 & 3.3  & M2 & & \citet{Torres2006}  \\
V1086 Tau &  & 1.0 & 0.09 & 11.1 & K7 & & \citet{Herbig1988}  \\
V396 Aur  &  & 2.4 & 0.35 & 6.7  & K0 & & \citet{Herbig1988}  \\
V397 Aur  &  & 1.7 & 0.26 & 6.3  & K7 & & \citet{Steffen2001} \\

\hline
\multicolumn{8}{l}{$^1$The 1.3~mm flux has been renormalized to mJy at a distance of 140 pc.}\\
\multicolumn{8}{l}{$^2$The Si:N, Si:C and C:N hot gas line flux ratios are from \citet{Ardila2013}.}\\
\multicolumn{8}{l}{$^3$The 1.3~mm flux has been extrapolated from the 880$\umu m$ flux.}\\
\multicolumn{8}{l}{$^4$The 1.3~mm flux has been estimated by extrapolating the far infra-red SED.}\\
\multicolumn{8}{l}{$^5$The line profiles of the sources marked `Yes' may be affected by outflows.}
\end{tabular}
\end{table*}

\begin{figure*}
\includegraphics[width=\textwidth]{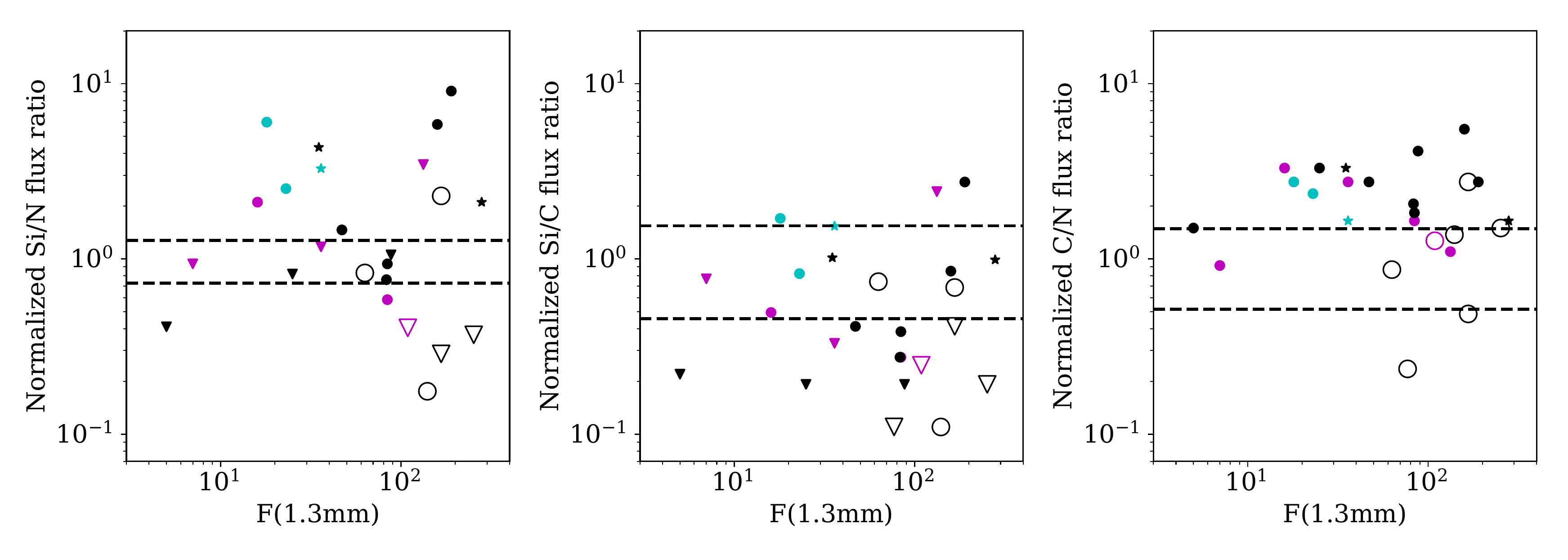}
\caption{Flux ratios of Si, N and C normalized to the WTT sample mean against the flux at 1.3~mm normalized to the distance of Taurus  (140pc). The 1-$\sigma$ deviation of the WTT sample are shown by the black dashed lines. Filled circles denote full discs, while large hollow symbols refer to sources with known large cavities. Triangles denote upper limits. The sources with line profiles that may be affected by outflows are marked with stars. The spectral type of the sources is shown by point colour: M-stars are in magenta, F and G stars in cyan and K stars in black. Note that for the cavity source V 4046 Sgr the normalized Si/N ratio ( 0.03)  lies below  the bottom of the figure.}
\label{Fig:ObsCNSi}
\end{figure*}

  We consider  a sample of  stars with spectral types F5-M2 from those analysed by \citet{Ardila2013}  comprising 25 stars where there is  evidence of circumstellar emission and
5 Weak Line T Tauri stars (WTTs). Among the former sample we define  7 `cavity
sources' as those for which there is evidence from imaging of a cavity in mm
emission (these sources overlap with those
denoted `transition discs' by \citealt{Ardila2013}). The sources, along with their classification are listed  in \autoref{Tab:Data}.
 Given that not all sources have been examined with the
same resolution and sensitivity, we note that these sources represent a lower limit
to those with cavities and we cannot rule out that there are
additional cavity sources concealed in the data (see below). 

  We follow  \citet{Ardila2013}  in considering the {\it ratios} of line fluxes, rather
than their absolute values, because the data has not been corrected for extinction.
It is therefore implicit in our analysis that the emission regions of the
various species are co-spatial and subject to the same extinction. This is probably
justified under the paradigm that we are detecting emission from localised
regions such as accretion columns or upper regions of the stellar atmosphere. In considering
line ratios indicative of the abundances of N, C and Si we sum luminosities in the \NV{} lines at $1238.88\,\angstrom$ and $1242.88\,\angstrom$, in \CIV{} the doublet at $1548.28\,\angstrom$ and 
$1550.8$, and in \SiIV{} the lines at $1393.8\,\angstrom$ and $1402.88\,\angstrom$.

 \citet{Ardila2013}  discussed the luminosity ratios of the WTTs in their sample in terms of a model in which the lines are assumed to be optically thin, in a state of collisional equilibrium and deriving from a region in the upper stellar atmosphere (the transition region) with a
 single temperature and ionisation parameter (optically thin emission is suggested by the mean ratio of the components of the \CIV{} doublet: see \citet{Ardila2013}  for a critical discussion of these
 assumptions). They  found that when applied to the C/N luminosity ratios in the WTT sample, the 
 measured ratios were consistent with solar abundances, as expected.
 In the case of line ratios involving Si, they instead  noted the anomalously bright Si emission compared with model predictions, but that a similar enhancement is seen in the solar spectrum. This latter phenomenon, termed the First Ionisation Potential  (FIP) effect, is not fully understood  \citep{Vauclair1985,Laming2004} but the magnitude
 of the excess (around a factor three excess in the Si luminosity) is consistent between the WTT sample and the Sun.  The mean line ratios observed in the WTT sample were thus found to be consistent with those expected for solar abundances.\footnote{Note that the use of data that has not been corrected for extinction introduces an element of scatter
on account of the steep rise of the extinction curve in the wavelength
region producing the three diagnostic lines. For example an extinction
of $A_v =1.0$ mag leads to a change in flux ratio of $2.4$ between
the lines that are most separated in wavelength \citep[e.g.][]{Mathis1990}. This effect will add some scatter
to the flux ratios for both WTTs and CTTs. We however doubt that this
is the dominant source of scatter in either sample because there is no evidence that
the scatter is larger for the line ratios involving the largest
difference in wavelength (i.e. C/N).}
 
 When considering the line ratios in the CTT sample there are many effects that may change the line ratios without necessarily implying a change in abundances. For example, in collisional equilibrium the emissivity of \SiIV{} peaks at much lower temperature than the other species; a difference in the temperature of emitting regions between CTTs and WTTs, as well as  variations in both temperature and radiation field among the CTT population,  could therefore give rise to variations in line fluxes which are not primarily driven by changes in abundance. For example the much larger {\it scatter} in line ratios in the CTT case could plausibly be attributed to the fact that in this case the lines are generated in multiple regions, with contributions from outflows, stellar magnetic activity and accretion. Even in the case where the lines are predominantly associated with accretion, there are likely to be contributions from multiple components (the pre-shocked gas and the heated stellar photosphere) which are likely to have different temperatures and radiation fields that can vary between sources (\citealt{Beristain2001,Yang2007,Ingleby2013}; see also the review by \citealt{Hartmann2016}). In the case of multiple components, the total line fluxes then depend on their relative dominance  which could depend on factors such as inclination  (however \citealt{Ardila2013} found little evidence for strong inclination dependence of the line fluxes). In \autoref{Fig:ObsCNSi} we code the data points according to stellar spectral type and indicate whether there is evidence for an outflow that could affect the shape of the line profiles. While not being in any sense conclusive we do not see obvious evidence from this plot that the line ratios are systematically correlated with photospheric temperature or with outflow characteristics. However, the other factors likely contribute at the very least to the scatter.
 
 In this paper we proceed by examining the intriguing possibility that the major source of the scatter in line ratios among CTTs reflects differences in atomic abundances in the accretion flow. In Section 3 we  will examine whether the implied abundance variations and possible trends with disc parameters  could in principle be understood  in terms of current models in which the transport of chemical species on radially migrating dust grains plays a major role. Although we note that some of the known uncertainties (such as the magnitude of the FIP effect) introduces an uncertainty that is small compared with the total spread in line ratios seen in the  CTT sample, we are mindful that the uncertainties discussed above, relating to the differences in temperature and radiative regime, are much harder to quantify:
 these uncertainties affect not only the interpretation of the scatter in
 line ratios among the CTT sample but also the comparison between the CTT and WTT samples. Thus although we will interpret the large scatter of line ratios in CTTs in terms of  abundance variations, we  remain  mindful that there are a number of other effects that  may contribute to this scatter.

  In \autoref{Fig:ObsCNSi} we  plot the line ratios of the CTT sample normalised to the mean of the WTT sample, with the dashed lines representing the mean $\pm 1 \sigma$ of the WTT sample. In the most naive interpretation, where variations in line ratios are a direct measure of elemental abundance, the normalised flux ratios would therefore represent abundance ratios normalised to solar. While this association is uncertain for a variety of reasons, including the possibility of different temperature distributions between the WTT and CTT sample, \citet{Ardila2013} show the C/N fluxes are in reasonable agreement across the two samples, suggesting the WTT line ratios may be a reasonable benchmark. The flux ratios are plotted against the mm flux: this choice is motivated by the fact that the driver of variations of abundance ratio in the models presented in Section 3 is the growth and radial migration of ice covered grains and that this  drives  changes in the mm flux through its impact on the size of the mm emitting disc and the grain opacity. 
  
 From \autoref{Fig:ObsCNSi} it is immediately obvious that the scatter in flux ratios is very high in the CTT sample compared with the
 WTT sample and we will first examine whether there is any circumstantial evidence to connect this
 variation with abundance variations. The most obvious indication that abundance variations are playing a role
 is the lower values of Si/N and C/N luminosity ratios of the sources with a cavity imaged in the mm (with the logrank Kaplan-Meier test \citep{Feigelson1985} indicating that the cavity and non-cavity sources differ in their distribution of Si/N and C/N luminosity ratios at a 2.4 and $2.9 \sigma$ level respectively). Such an outcome is expected if the cavity involves the
 preferential retention of Si at large radii in the disc due to trapping of large particles at the cavity rim. Similarly, the low C/N ratio of the cavity sources can be explained simply because the cavity rims typically fall between the CO$_2$ and CO or N$_2$ snow lines (a few, and a few 10$\unit{au}$, respectively). This would imply that the nitrogen abundance is dominated by highly volatile species such as N$_2$, which should be in the gas phase at the cavity rims, while a significant fraction of carbon is found in molecules such as CO$_2$ or refractory organics, which may remain at the rim trapped in ices, rather than being in the form of volatile CO.  Further support is provided by the connection between the spectral energy distributions (SEDs) of the
 cavity sources and their location in this diagram: objects with `clean cavities' (weak or absent NIR excess) are located
 at low values of normalised flux ratios, while the two sources with higher Si/N line ratios  (Lk Ca 15 and UX Tau A) have SEDs characteristic of pre-transition discs with an optically thick inner disc \citep{Espaillat2007}. 
 
 Excluding the cavity sources, we see that the Si/N normalised luminosity ratio is positively correlated
 with mm flux (with coefficient 0.35, only consistent with zero correlation at the 10 per cent level). Given the evidence that mm flux declines with time \citep{Barenfeld2016,Pascucci2016,Ansdell2016} (plausibly on account of the radial migration process modeled in Section 3 which drives a reduction in both the dust mass and the size of the mm emitting disc: see e.g. \citealt{Pietu2014,vanderPlas2016,Hendler2017}),  this gives some suggestion that, in the absence of a mm cavity, the Si content of accreting material likewise declines with age. The middle panel depicts a similar correlation for Si/C though with more scatter (also only consistent with zero correlation at the 10 per cent level). The right hand panel shows no obvious relationship between C/N luminosity ratio and mm flux. Nevertheless it notable that the bulk of the sample has a normalised C/N luminosity ratios that is a factor few higher than that in the WTTs. 
  
In the following Section we will interpret these plots under the extreme assumption that they
are largely driven by abundance variations. In this paradigm we see that the largest abundance variations are seen in relation to Si and N while C occupies an intermediate position. In outline, we will seek to understand this behaviour in terms of the relative importance of transport on grains and in the gas phase for the three elements concerned, arguing that the ordering of the ice lines for the dominant carriers implies an ordering of the arrival times of these species into the innermost disc. We however emphasise the caveats listed above concerning alternative interpretations of the line ratio data.

\section{Models}
\label{Sec:Model}

\begin{table*}
  \centering
 \begin{tabular}{cccc}
  \hline \hline
  Species (X) & T$_{\rm bind}$ [K]$^a$ & Case 1$^b$: X/H & Case 2$^c$: X/H \\ \hline
  N$_2$  &  770 & $0.5\times ($N/H$ - $NH$_3$/H) & $0.5\times ($N/H$ - $NH$_3$/H) \\
  CO     &  850 & 0.45 ($1 + f_{{\rm CO}_2}$) $\times$ C/H   & 0.65 $\times$ C/H \\
  CH$_4$ & 1300 & 0.45 ($1 - f_{{\rm CO}_2}$) $\times$ C/H   & 0 \\
  CO$_2$ & 2575 & 0.1  $\times$ C/H           & 0.15 $\times$ C/H \\
  NH$_3$ & 2965 & 0.07 $\times$ N/H           & 0.07 $\times$ N/H \\
  H$_2$O & 5700 & O/H - (3 $\times$ Si/H + CO/H + 2 $\times$ CO$_2$/H) & O/H - (3 $\times$ Si/H + CO/H + 2 $\times$ CO$_2$/H) \\
  Carbon grains & -- & 0 & 0.2 $\times$ C/H \\
  Silicates     & -- & Si/H & Si/H \\
  \hline
  \end{tabular}
  \caption{Binding energies and volume mixing ratios of chemical species adopted \citep[from][]{Madhusudhan2014c,Booth2017}.}
  \label{Tab:Chem}
  {$^a$ We use binding energies \citet{Collings2004,Garrod2006,Martin-Domenech2014} and \citet{Fayolle2016}.\\
  $^{b,c}$ Volume mixing ratios adopted from \citet[see e.g.][]{Madhusudhan2014c}, with solar elemental abundances \citep{Asplund2009}. $f_{{\rm CO}_2}$ is the fraction of CO$_2$ molecules in ices. The NH$_3$ abundance of 0.07 follows \citet{Bottinelli2010} and \citet{Piso2016}.}
\end{table*}

\begin{figure*}
\includegraphics[width=\textwidth]{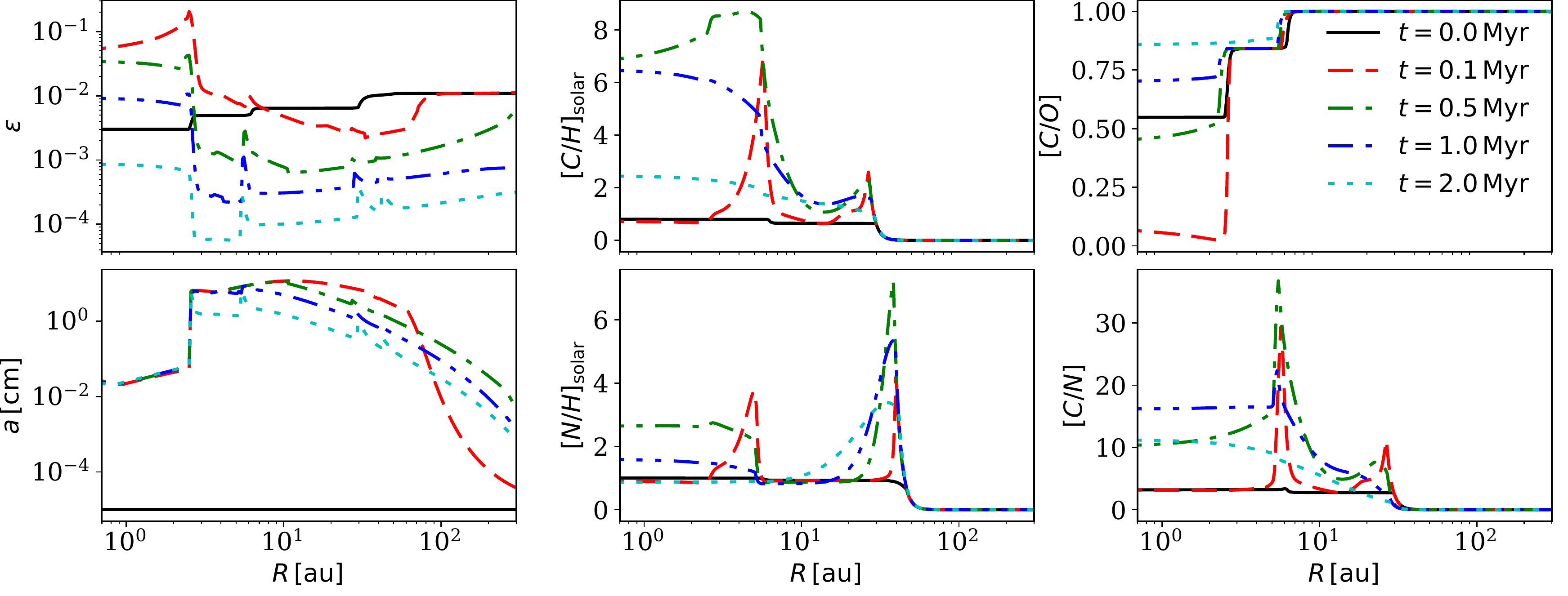}
\caption{Evolution of the dust fraction, $\epsilon$, maximum grain size, $a$, and gas phase chemical abundances for a 100\unit{au} disc. The C/H and N/H ratios are normalized to the solar values.}
\label{Fig:ModelEvol}
\end{figure*}

\begin{figure*}
\includegraphics[width=\textwidth]{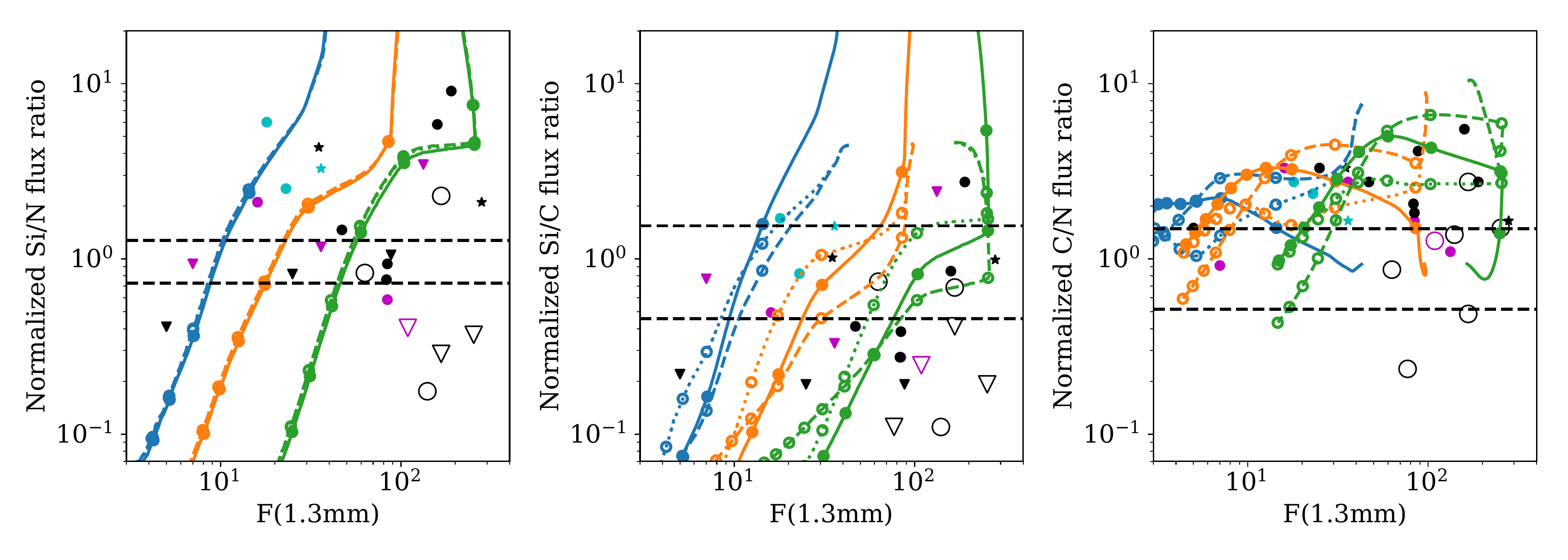}
\caption{Same as \autoref{Fig:ObsCNSi}, with the addition of model tracks that show ratio of the atomic species accreted onto star with dots placed every 0.5\unit{Myr}. The line colours denote the initial disc outer radii: 50\unit{au} (blue), 100\unit{au} (orange), and 300\unit{au} (green). The Case 1 and 2 `equilibrium' models are shown by solid and dashed lines, while the dotted line shows that Case 2 `no reactions' model.}
\label{Fig:ModelCNSi}
\end{figure*}

We follow the evolution of the chemical species accreting onto the star using the disc model of \citet{Booth2017}. This model treats grain growth and radial drift following \citet{Birnstiel2012}, along with gas evolution due to turbulent viscosity and diffusion. The mid-plane temperature is computed including viscous heating and irradiation. The stellar parameters  and  disc initial conditions are the same values as \citet{Booth2017}, with an initial accretion rate of  $10^{-8}M_\odot\unit{yr}^{-1}$ and a viscous $\alpha$ parameter of $10^{-3}$. 

The chemical model follows the transport of molecular species due to radial drift, viscous accretion and diffusion of ices and vapours. For the chemical abundances we use the models presented in \citet{Booth2017}, which include the adsorption and desorption of molecular species. Here we also include N$_2$ and NH$_3$, the dominant nitrogen bearing species following \citet{Piso2016}. The abundances of the molecular species and their binding energies are given in \autoref{Tab:Chem}. As in \citet{Booth2017} we explore the importance of chemical reactions by studying the two extremes: neglecting them entirely or assuming that they are fast so that  C bearing species are driven to an `equilibrium' given by \autoref{Tab:Chem}. Since chemical reactions only significantly affect N bearing species close to and inside the NH$_3$ snow line \citep{Eistrup2016} they do not affect the transport of nitrogen and thus we always neglect them.

This simplified chemical model neglects a number effects associated with the full complexity of chemistry in protoplanetary discs \citep[see, e.g.,][for a review]{PontoppidanPPVI}, e.g. the possibility of carbon being locked up into complex organics. However, we will argue that, based on the work presented here and in \citet{Kama2015}, the carbon-locking is not the main driver for the chemical evolution of the inner regions of protoplanetary discs. As discussed in \citet{Booth2017}, in the absence of detailed chemical models that include radial drift, we gain some confidence in the simple models from the fact that they span the range of behaviours seen in the detailed chemical models of \citet{Eistrup2016}.

\autoref{Fig:ModelEvol} shows the typical evolution, using a disc with an initial mass  of $0.1\,M_\odot$, radius of 100\unit{au}, and Case 2 `equilibrium' chemistry. The physical evolution is typical for models of this kind, with the mass of the gas decreasing due to viscous accretion, while it spreads outwards in radius. In comparison, the dust mass evolves more quickly due to growth and radial drift, leading to the decrease of both the dust-to-gas ratio and disc size (as determined by, e.g. the radius outside of which the dust is smaller than 1\unit{mm}). 

The dust evolution is associated with the transport of volatiles that make up the icy  mantles of the grains, which are carried as the dust migrates. These ices are deposited by desorption at their respective snow lines while the grains continue to drift towards the star. CO and N$_2$ are left furthest out, at 30--40\unit{au}, leading to spikes in the C/H and N/H ratios. Next CO$_2$ and NH$_3$ are left at 5--6\unit{au}, with water finally desorbing at around 2\unit{au}. 

This evolution results in a partial sorting of elemental abundances based on where most of the mass is deposited. Si, being present only in dust grains is deposited closest to the star. Next is C, with N being deposited further out. The C and N profiles are each controlled by two important snow-lines: CO and N$_2$ at large radius and CO$_2$ and NH$_3$ closer in: however the  fraction of C residing  in CO$_2$ is large compared with the fraction of N in NH$_3$ and thus the net effect is that carbon is more rapidly conveyed to the inner disc than N.  In \autoref{Fig:ModelEvol} the difference between C and N is further enhanced because the conversion of CO to CO$_2$ at the snow line results in extra C being deposited at the CO$_2$ snow line. 

We now show this sorting of atomic abundances can explain the main feature of the \citet{Ardila2013} data if one interprets the flux ratios as indicating  the abundance ratios. We note that the abundance evolution of the accretion flow  can already be inferred from \autoref{Fig:ModelEvol} because the abundance of accreting material tracks that  at the inner edge of model. This is also true of the dust, which is well coupled to the gas at the inner edge due to its small size (thus the dust-to-gas ratio normalized to its initial value gives the Si abundance relative to solar). The 1.3\unit{mm} flux for the models is derived assuming a \citet{Weingartner2001} composition. The grain size distribution is $n(a)\diff a \propto a^{-q} \diff a$, with $q=3$ (appropriate for an evolved dust population, \citealt{Birnstiel2011}) and the maximum grain size determined by the model. The flux is computed using the temperatures from the disc model, assuming a face-on disc and including the effects of optical depth.

Three models with outer radii of 50, 100, and 300\unit{au} are plotted along with the data in \autoref{Fig:ModelCNSi}, including results for both of the `equilibrium' chemistry models and the Case 2  `no reactions' chemistry. The models do a remarkable job of explaining the parameter space filled by the non-cavity sources. In particular, they capture the correlation seen in the Si/N data, i.e. young or massive discs have high Si/N and mm-flux which both decrease as the discs evolve. The models also explain the absence  of non-cavity sources in the upper left of the Si/N -- mm flux diagram. Although discs with small initial radii ($\sim 10\unit{au}$) would fill this parameter space, they evolve faster (which can be seen by the points, starting at 0.5\unit{Myr} and spaced every 0.5\unit{Myr} thereafter) and have left this region within a Myr. 

The ages of model discs with high Si/N ratios suggest that these sources should be young, $\sim 0.5\,{\rm Myr}$, while sources with lower Si/N should be considerably older. While this age trend is a fundamental prediction of the model, we do not place any weight on the particular ages themselves, due to both the simplicity of the model itself, and also because an offset in the normalization of the flux ratio when interpreting them as abundances would manifest itself as a systematic offset in the age.

Considering now the Si/C and C/N data, we again see that the models successfully explore the parameter space filled by the data. The super-solar C/N ratio in the models is paralleled by the line flux ratios; however, we note that due to the possibility of different temperatures in the emission regions of WTTs compared with CTTs leading to an offset in the zero points, this association must be taken with caution. A possible sign of consistency is that the disc size implied by the data points in Si/N is roughly consistent with the one implied by the Si/C (e.g. the discs that fall between the 50 and 100\unit{au} disc radii for Si/N are in a similar place in the Si/C plot). 

Finally, the low Si abundance of the cavity sources can be explained by a dust trap that stops the flow of dust into the inner disc, resulting in a low Si abundance (excepting the sources with a significant NIR excess). The lower C/N ratio of the cavity sources also suggests that more carbon is being trapped in the grains than nitrogen. Given the radii of these cavities are comparable to, or beyond, the CO$_2$ snow line, this could be either in the form of CO$_2$ ice or complex organics. We note that whereas {\it all} the known cavity sources in the sample lie to the lower right of the model trajectory for an initial disc radius of
300 \unit{au}, there are a small number of non-cavity sources which occupy a similar region of
parameter space in the mm flux, Si/N plane. In particular the cluster of non cavity sources with high mm flux and relatively low Si/N emission ratios (MP Mus, AA Tau, DN Tau and EP Cha) are all likely candidates for possession of a mm cavity on this basis and it will be
interesting to see whether further ALMA imaging of these discs bears out this prediction. 

\section{Discussion \& Conclusion}

We have re-examined the UV line ratio dataset from \citet{Ardila2013} in the context of a modern understanding of the role of dust evolution on the chemical evolution of protoplanetary discs. In this picture, chemical evolution is driven by the flux of volatiles carried in the icy mantels of large dust grains as they drift rapidly towards the star. Since these volatiles leave the icy mantels and re-enter the gas phase at snow lines, while the dust grains continue to drift, this gives rise to a partial sorting of the elemental abundances based upon the volatility of the dominant molecular carrier. Thus, the disc becomes more depleted in dust than nitrogen, with carbon being intermediate. We show that this leads to the Si/N ratio being correlated with mm-flux and predict an enhanced C/N ratio in the inner disc. 

Under the crude assumption that the variation in the UV line ratio data from \citet{Ardila2013} is indicative of variation in the elemental abundances, the data is in remarkable agreement with the models. In particular, the data for CTTS without imaged mm cavities show a similar correlation between the Si/N line ratio and 1.3~mm flux. Furthermore, if one makes the somewhat dubious assumption that the elemental ratios can be determined by comparing them to the WTTs line ratios (which should be typical of the stellar abundance, but may have a different temperature distribution), then the CTTs also show an enhanced C/N ratio that is compatible with the models. If the C/N ratio is indeed enhanced, then this is hard to reconcile with he origin of CO depletion in protoplanetary discs being due to the vast majority of C being locked into refractory complex organics, unless N is also locked into the organics at a similar, but slightly lower level.

However, we note that the interpretation of line ratios as direct tracers of elemental abundance is challenging since many effects (including different temperature distributions in the emitting region, along with other effects discussed in Section 2 and \citealt{Ardila2013}), may cause significant variations in line ratios. In spite of this, the fact that discs with large mm cavities show systematically lower Si/N emission than full discs gives us some confidence that abundance variations are indeed playing a role. This outcome is naturally explained if large dust grains are being trapped at the cavity rim, which has already been seen in Herbig stars where the abundance determinations are more robust \citep{Folsom2012,Kama2015} and fits into our picture chemical evolution driven by dust evolution. In this interpretation, the lower C/N ratios of the cavity sources suggests that more nitrogen than carbon should be in the gas phase at the cavity edges, which should be the case if most of the nitrogen is the form of N$_2$ (as we assume here) and the cavity edges fall between the CO and CO$_2$ snow lines. Thus, we argue that a robust link between line ratios and abundance ratios, along with larger samples, has the potential to cast considerable light on the chemical history of protoplanetary discs.

\section*{Acknowledgements}

This work has been supported by the DISCSIM project, grant agreement 341137 funded by the European Research Council under ERC-2013-ADG. We thank Mihkel Kama and James Owen for many fruitful discussions. We are indebted to the referee, Greg Herczeg,  for many insightful comments. This work was partly developed during and benefited from the MIAPP ``Protoplanetary discs and planet formation and evolution'' program.

\bibliography{chemo_drift}
\bibliographystyle{mnras_edit}

\bsp

\label{lastpage}

\end{document}